# GitHub Marketplace: Driving Automation and Fostering Innovation in Software Development

SK. Golam Saroar, Waseefa Ahmed, Elmira Onagh, Maleknaz Nayebi
EXINES Lab, York University, Canada
Email: {saroar, waseefa, eonagh, mnayebi}@yorku.ca
**Journal First Paper from Information and Software Technology (IST)**

## I. EXTENDED ABSTRACT

GitHub, a central hub for collaborative software development, has revolutionized the open-source software (OSS) ecosystem through its GitHub Marketplace, a platform launched in 2017 to host automation tools aimed at enhancing the efficiency and scalability of software projects. As the adoption of automation in OSS production grows, understanding the trends, characteristics, and underlying dynamics of this marketplace has become vital. Furthermore, despite the rich repository of academic research on software automation, a disconnect persists between academia and industry practices. This study seeks to bridge this gap by providing a systematic analysis of the GitHub Marketplace, comparing trends observed in industry tools with advancements reported in academic literature, and identifying areas where academia can contribute to practical innovation.

**Introduction.** The rapid evolution of software development has driven the need for efficient automation tools that simplify tasks such as continuous integration, testing, and deployment. GitHub Marketplace, featuring 8,318 automation tools (440 Apps and 7,878 Actions), offers a unique opportunity to explore the state-of-practice in OSS automation. Simultaneously, the academic community has made significant strides in software automation research, contributing novel algorithms and tools that address critical challenges in OSS production. However, adopting academic tools in practice remains limited, raising questions about the alignment between research efforts and industry needs.

**Objective.** This study addresses two fundamental questions:
1) What trends and characteristics define the GitHub Marketplace and the tools hosted on it?
2) How do these trends compare to advancements in academic research on OSS automation, and where do gaps exist?

Our goal is to provide actionable insights that help researchers align their work with industry demands while enabling practitioners to better understand and evaluate the tools available.

**Methodology.** To address the research questions, we employed a two-pronged approach. First, an industry analysis was conducted using data from the GitHub Marketplace, where we categorized tools into 32 functional types. This analysis focused on adoption trends, tool ratings, and the influence of verification in building user trust. Second, an academic analysis involved a systematic literature review of 515 research papers published between 2000 and 2021, examining OSS automation contributions, with particular emphasis on tools developed for testing, continuous integration, and code quality. This combined methodology allowed for a comprehensive examination of industry and academic contributions, facilitating a side-by-side comparison of their impacts and advancements.

**Results and Discussion.** Our analysis revealed several important findings. First, there is a divergence in priorities between industry and academia. The industry tends to prioritize tools for "Continuous Integration" (CI) and "Utilities," reflecting the demand for streamlined workflows in collaborative development environments. In contrast, academic research focuses more on "Code Quality" and "Testing," with limited attention given to CI and deployment tools, which are critical in practical settings. Second, the role of trust and verification emerged as a key factor in tool adoption. Tools that are verified by GitHub or associated with verified domains are significantly more likely to gain user trust and adoption, suggesting that academic tools, which are often under-adopted, could benefit from partnerships with trusted entities or platforms to enhance their visibility and credibility. Lastly, our findings highlight a limited translation of academic tools into widespread practice.

Despite decades of research, many academic tools fail to gain traction in the industry due to issues such as lack of accessibility, insufficient documentation, and misalignment with real-world needs and requirements.

**Conclusion.** In conclusion, this study highlights the critical need for stronger collaboration between academia and industry to advance OSS automation. By prioritizing areas such as DevOps and continuous deployment, researchers can align their efforts with industry needs while addressing barriers to adoption through improved usability, documentation, and partnerships. Industry practitioners, in turn, can benefit from academic insights to drive tool development and innovation. Future research should focus on mining repositories for deeper insights into adoption patterns, addressing gaps in tool functionality, and establishing standardized benchmarks for tool evaluation. By bridging the divide between academic research and industry practices, this study provides a roadmap for fostering collaboration, addressing real-world challenges, and driving innovation in software engineering. Such alignment is essential for pushing the boundaries of OSS automation and ensuring impactful contributions to both fields.

## II. Paper Information

- **Paper title:** GitHub marketplace for automation and innovation in software production
- **Authors:** SK. Golam Saroar, Waseefa Ahmed, Elmira Onagh, Maleknaz Nayebi
- **Acceptance date:** July 3, 2024
- **Publication date:** Available online 14 July 2024, published in Nov 2024 special issue.
- **pointer to original paper:**
  https://www.sciencedirect.com/science/article/pii/S0950584924001277

## III. J1C2 criteria satisfaction

This paper meets the J1C2 criteria for the SANER 2025 conference as it presents new research results, offering a systematic analysis of the GitHub Marketplace and comparing industry trends with academic advancements in OSS automation. It explores software evolution and maintenance by analyzing over 8,000 tools in the Marketplace, highlighting gaps between academic research and industry practices. The work is original, not an extension of any prior conference publication, and was accepted and published in the *Information and Software Technology journal* (published online July 14, 2024). It has not been presented in any other journal-first program, ensuring full adherence to J1C2 guidelines. This study is highly relevant to the SANER community, specifically in the areas of Software Tools for Software Evolution and Maintenance and Mining Software Repositories and Software Analytics, as it provides valuable insights into industry trends and the adoption of automation tools in software engineering.

## IV. Confirmation of New Research Results

This paper presents new research results and is not an extension of any previous conference publication. It provides an original analysis of the GitHub Marketplace, comparing industry trends with academic advancements in OSS automation, and offers new insights into the adoption patterns and challenges faced by academic tools in the industry. The work is based on a comprehensive analysis of GitHub Marketplace tools and academic literature, presenting findings that have not been published in any prior conference proceedings.

## V. Confirmation of No Prior Journal-First Program Presentation

This paper has not been presented at or is currently under consideration for any journal-first program of another conference, nor has it been presented in any past edition of this conference. **The work is original and exclusively submitted for the J1C2 track of SANER 2025. This work is the result of long-term research on software ecosystems, drawing on numerous studies, as referenced in [9].**

This work builds upon the existing body of literature developed over the years [6], [1], [37], [39], [5], [11], [10], [3], [15], [31], [8], [9], [38], [34], [2], [22], [21], [14], [23], [26], [18], [12], [13], [29], [7], [20], [28], [35], [30], [25], [17], [19], [36], [24], [16], [4], [27], [2], [26], [40], [41], [33], [9], [38], [40], [41], [33], [32].